\documentclass[preprint,showpacs,preprintnumbers,amsmath,amssymb]{revtex4-1}

\usepackage{graphicx}
\usepackage{color}
\usepackage{comment}
\usepackage[normalem]{ulem}

\renewcommand{\bm}[1]{{\mbox{\boldmath $#1$}}}
\newcommand{\FST}{FeSe$_{1-x}$Te$_{x}$}
\newcommand{\FSS}{FeSe$_{1-x}$S$_{x}$}
\newcommand{\msr}{$\mu$\text{SR}}

\begin{document}
\title{Topology meets time-reversal symmetry breaking in FeSe$\bm{_{1-x}}$Te$\bm{_{x}}$ superconductors}
\author{M.~Roppongi$^1$}\email{roppongi@qpm.k.u-tokyo.ac.jp}
\author{Y.~Cai$^{2,3}$}
\author{K.~Ogawa$^1$}
\author{S.~Liu$^1$}
\author{G.~Q.~Zhao$^4$}
\author{M.~Oudah$^{2,3}$}
\author{T.~Fujii$^5$}
\author{K.~Imamura$^1$}
\author{S.~Fang$^1$}
\author{K.~Ishihara$^1$}
\author{K.~Hashimoto$^1$}
\author{K.~Matsuura$^6$}
\author{Y.~Mizukami$^7$}
\author{M.~Pula$^8$}
\author{C.~Young$^3$}
\author{I.~Markovi\'{c}$^{2,3}$}
\author{D.~A.~Bonn$^{2,3}$}
\author{T.~Watanabe$^9$}
\author{A.~Yamashita$^{10}$}
\author{Y.~Mizuguchi$^{10}$}
\author{G.~M.~Luke$^8$}
\author{K.~M.~Kojima$^{2,3}$}
\author{Y.~J.~Uemura$^{11}$}\email{yu2@columbia.edu}
\author{T.~Shibauchi$^1$}\email{shibauchi@k.u-tokyo.ac.jp}

\affiliation{$^1$Department of Advanced Materials Science, University of Tokyo, Kashiwa, Chiba 277-8561, Japan}
\affiliation{$^2$Stewart Blusson Quantum Matter Institute, University of British Columbia, Vancouver, British Columbia V6T 1Z4}
\affiliation{$^3$Department of Physics and Astronomy, University of British Columbia, Vancouver, Canada V6T 1Z1}
\affiliation{$^4$Kavli Institute for Theoretical Sciences, University of Chinese Academy of Sciences, Beijing, 101408, China}
\affiliation{$^5$Cryogenic Research Center, University of Tokyo, Bunkyo-ku, Tokyo 113-0032, Japan}
\affiliation{$^6$Department of Applied Physics, University of Tokyo, Bunkyo-ku, Tokyo 113-8656, Japan}
\affiliation{$^7$Department of Physics, Tohoku University, Sendai 980-8578, Japan}
\affiliation{$^8$Department of Physics and Astronomy, McMaster University, Hamilton, Ontario, Canada, L8S 4M1}
\affiliation{$^9$Graduate School of Science and Technology, Hirosaki University, Hirosaki, Aomori 036-8561, Japan}
\affiliation{$^{10}$Department of Physics, Tokyo Metropolitan University, Hachioji 192-0397, Japan}
\affiliation{$^{11}$Department of Physics, Columbia University, New York, NY 10027}
\date{\today}

\begin{abstract}
{\bf 
Time-reversal symmetry breaking (TRSB) in magnetic topological insulators induces a Dirac gap in the topological surface state (TSS), leading to exotic phenomena such as the quantum anomalous Hall effect. Yet, the interplay between TRSB and topology in superconductors remains underexplored due to limited suitable materials. Here we employ zero-field muon spin relaxation ($\bm{\mu}$SR) as a sensitive probe of TRSB to map out the electronic phase diagrams of iron-chalcogenide superconductors FeSe$\bm{_{1-x}}$Te$\bm{_{x}}$. For the Te composition $\bm{x=0.64}$ with the highest superconducting transition temperature $T_{\rm c}=14.5$\,K, which is known to host a TSS and Majorana zero modes within vortices, we detect spontaneous magnetic fields below $T_{\rm c}$ distinct from a magnetic order. This signifies a TRSB superconducting state in the bulk, revealing the convergence of unconventional TRSB superconductivity with topologically nontrivial electronic structures in FeSe$\bm{_{1-x}}$Te$\bm{_{x}}$. Given the relatively high $T_{\rm c}$ and the tunability of the Fermi level through chemical substitution, iron-chalcogenide superconductors offer an intriguing platform for investigating the synergy between topological superconductivity and TRSB. 
}
\end{abstract}
\clearpage
\maketitle

\begin{figure*}[tbp]
    \includegraphics[width=0.8\linewidth]{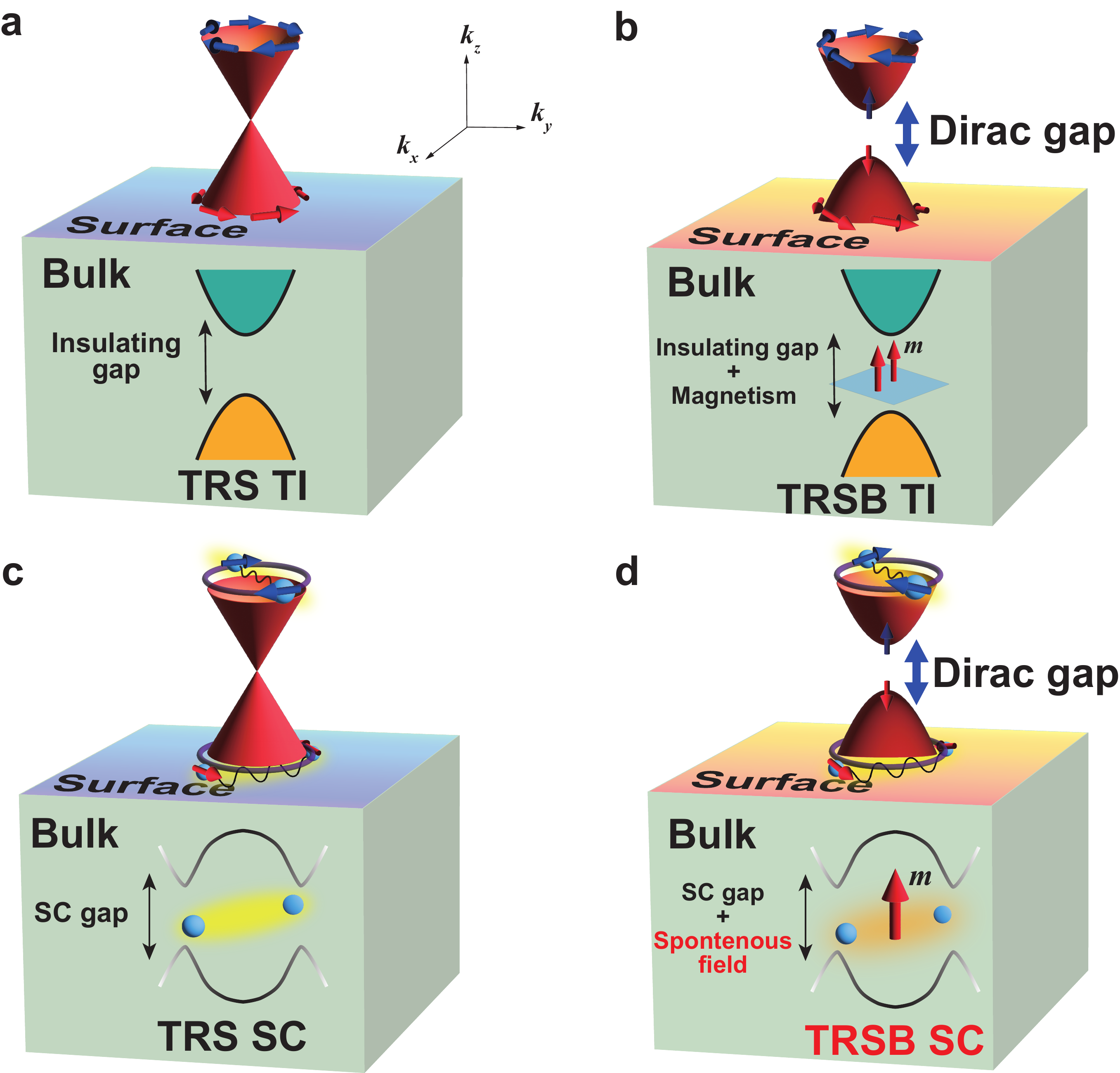}
    \caption{{\bf Schematic comparisons between topological insulators (TIs) and superconductors (SCs) having topological surface states (TSSs) with and without time-reversal symmetry (TRS).} {\bf a}, In TIs with TRS, the insulating gap exists in the bulk, which closes at the surfaces with a Dirac-cone dispersion with spin momentum locking (red and blue arrows). {\bf b}, Magnetism in TIs breaks TRS, which induces the Dirac gap at the surfaces. {\bf c}, In TRS-preserved superconductors with topologically nontrivial electronic structures, the Dirac-like TSS will appear on the surfaces. {\bf d}, When TRS is broken in the bulk of superconductor with topological electronic structures, the spontaneous fields emerge in the bulk and the Dirac gap opens in the TSS. In {\bf c} and {\bf d}, the possible opening of the surface superconducting gap near the Fermi level due to proximity effect is not shown for simplicity. 
    }
\label{F1}
\end{figure*}

Symmetry breaking and topology are key concepts describing the electronic states in quantum materials. In topological insulators (TIs), the insulating band gap in the bulk is required to be closed at the surface, leading to the gapless Dirac-cone dispersion (Fig.\,\ref{F1}a), which is protected by time-reversal symmetry (TRS). In magnetic TIs, however, broken TRS leads to the opening of a gap in the TSS Dirac dispersion, as schematically shown in Fig.\,\ref{F1}b~\cite{Tokura2019, QAH_rev_2023}. With this Dirac gap, magnetic TIs with TRSB exhibit one-dimensional gapless chiral channels in the edges of top and bottom surfaces, leading to a quantum anomalous Hall insulator state or an axion insulator state depending on the absence or presence of inversion symmetry~\cite{Sekine2021}. 

Much less understood is the case for topological superconductivity. Enormous efforts have been devoted to the search for the Majorana bound states using several superconductors with topologically nontrivial order parameters or proximity-induced superconductivity in topological materials~\cite{Fu2008, Qi2011, Sato2016}, but the results and interpretations remain largely controversial. Another promising approach is to focus on bulk superconductors with topologically nontrivial electronic structures in the normal state~\cite{Sasaki2015,Neha2019,Zhang2018,Wang2018,Machida2019,Zhu2020,Ghosh2022,Shang2022}. Among others, the most significant Majorana evidence has been obtained in the iron-based superconductors \FST, which exhibit high $T_{\rm c}$ up to $\sim30$\,K under pressure~\cite{Mukasa2021} and high critical fields up to $\sim50$\,T~\cite{Mukasa2023}. In FeSe-based superconductors~\cite{Shibauchi2020}, the partial substitution of Se with Te shifts down the chalcogen $p$ band toward the Fermi level, making the topological band inversion along $\Gamma-Z$ direction of the Brillouin zone. Indeed, this results in the observation of a Dirac TSS in \FST\ ($x=0.55$)~\cite{Zhang2018, Li2024PRX}, and thus this system exhibits both bulk superconductivity and a topological electronic structure. Moreover, clear zero-bias conductance peaks in the fraction of superconducting vortices have been observed by the high-resolution scanning tunneling spectroscopy~\cite{Machida2019}, which can be reproduced by considering the Majorana bands and disordered vortex distribution~\cite{Chiu2020}. These results make \FST\ a strong candidate for topological superconductors. When TRS is broken in such a topological superconducting state, the Dirac gap is formed (Fig.\,\ref{F1}c,d), which can lead to exotic properties such as chiral edge modes and possibly non-Abelian statistics \cite{Qi2011}. 

To discuss the nature of superconductivity in \FST, let us start with the underlying general phase diagrams of the FeSe-based superconductors. Pure FeSe ($x = 0$) exhibits a nonmagnetic electronic nematic order, which can be suppressed by isovalent S or Te substitution for Se~\cite{Hosoi2016,Ishida2022}. In \FSS, recent \msr\ measurements reported evidence for TRSB in the superconducting state both in the nematic ($x<0.17$) and tetragonal ($x>0.17$) regimes~\cite{Matsuura2023}. It has been proposed that the TRSB superconductivity in the nematic phase of \FSS\ can be realized in an $s + {\rm e}^{{\rm i}\theta}d$-wave state with the phase $\theta$ changing from $\sim0$ to $\sim\pi$ near the nematic twin boundary~\cite{Matsuura2023,Kang2018,Watashige2015,Hashimoto2018}. In the tetragonal phase, an anomalously large residual density of states in the superconducting state has been observed~\cite{Matsuura2023,Hanaguri2018,Mizukami2023, Sato2018} in addition to TRSB, which is consistent with an ultranodal state with Bogoliubov Fermi surfaces~\cite{Setty2020}. In \FST, $T_{\rm c}$ initially decreases to a minimum ($T_{\rm c}\sim7$\,K) at $x$ $\sim0.3$, then rises again, reaching the maximum $T_{\rm c}\,\sim14.5$\,K near a nematic quantum critical point at $x \sim0.5$, where nematicity is completely suppressed~\cite{Mukasa2021, Ishida2022}. By combining the phase diagrams of \FSS\ and \FST, it has been discussed that these FeSe-based materials have three different superconducting (SC) phases; SC1 phase in tetragonal \FSS\ ($x>0.17$), SC2 phase showing the first dome connecting \FSS\ ($x<0.17$) and \FST\ ($x\lesssim0.3$), and SC3 phase showing the second dome in \FST\ ($x\gtrsim0.3$)~\cite{Ishida2022}. In this series, only the SC3 phase with the highest $T_{\rm c}$ maximum includes the topological region~\cite{Li2021}. As the TRSB superconductivity has been reported by \msr\ measurements in SC1 and SC2 phases, here we perform the \msr\ measurements for 5 different compositions in SC3 phase; $x\sim0.35(5)$ within the nematic regime, $x=$ 0.64(2), 0.75(1), 0.83(1), and 0.94(1) in the tetragonal regime (see Supplementary Information). We use collections of single crystals prepared by chemical vapor transport (CVT) ($x\sim0.35$) and by Bridgman method with Te-annealing ($x\ge 0.64$) (see Methods). 
It has been previously reported that low Te-substituted samples synthesized by CVT do not have excess irons\,\cite{Mukasa2021}.
In high-concentration samples synthesized by the Bridgman method, excess iron can be sufficiently removed by Te annealing\,\cite{Koshika2013, Watanabe2020, Fujii2023}. Indeed, in our measured samples, no significant effects of excess irons on the temperature dependence of magnetization and resistivity were observed at least for samples with $x\leq 0.75$ (see Supplementary Information).
Here we note that the end material FeTe ($x=1$) exhibits a long-range antiferromagnetic (AFM) order of double stripe-type, which suppresses bulk superconductivity~\cite{Tranquada2020}. Thus, the effect of spin fluctuations of this order should be considered for large $x$ regions. Moreover, recent experiments such as surface-sensitive photoemission and Kerr effect~\cite{Zaki2021, Li2021, Camron2023}, diamond magnetometry\,\cite{Mclaughlin2021},  as well as scanning superconducting quantum interference device (SQUID) microscopy and Josephson junctions using flakes~\cite{Lin2023, Qiu2023} have suggested the emergence of spontaneous ferromagnetism at the surfaces or interfaces of \FST\, in the topological region. To elucidate the nature of the superconducting state, it is essential to investigate the bulk state of \FST\ from the \msr\ measurements. 

\begin{figure*}[tbp]
    \includegraphics[width=\linewidth]{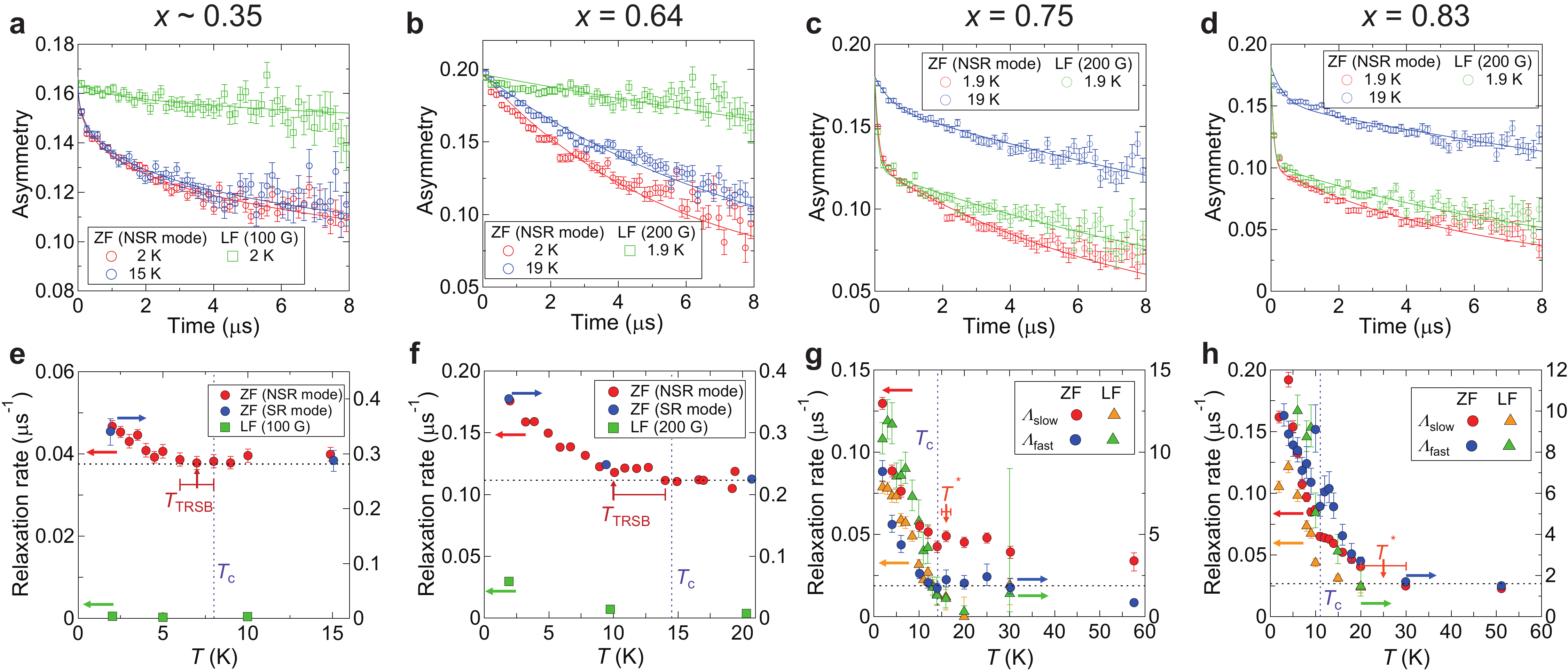}
    \caption{{\bf Zero-field (ZF) and longitudinal-field (LF) \msr\ in \FST\ superconductors.} {\bf a - d}, Time spectra of the \msr\ asymmetry in the non-spin rotated (NSR) mode for $x\sim 0.35$ ({\bf a}), $x=0.64$  ({\bf b}), $x=0.75$ ({\bf c}), and $x=0.83$ ({\bf d}). These spectra are representative of the lowest temperature below $T_c$ (ZF: red circles and LF: green squares), and above $T_c$ (blue circles (ZF)). Solid lines represent fitting curves. {\bf e - h}, Temperature dependence of relaxation rate for $x\sim 0.35$ ({\bf e}), $x=0.64$  ({\bf f}),  $x=0.75$ ({\bf g}), and $x=0.83$ ({\bf h}). The vertical dashed lines represent $T_{\rm c}$, and the onsets of TRSB ($T_{\rm TRSB}$) and short-range AFM ($T^*$) are indicated by arrows. Here, the relaxation rate of ZF NSR mode (red circle) and LF (red circles and green squares) correspond to the left axis, and those for ZF SR mode (blue circles) correspond to the right axis. For $x= 0.75$ ({\bf g}) and 0.83 ({\bf h}), the relaxation rate obtained by the fitting with the two-component relaxation function in ZF and LF-\msr\ (both in NSR mode) are displayed. The slow relaxation components (ZF: red circles and LF: yellow triangles) correspond to the left axis, while the fast relaxation components (ZF: blue circles and LF: green triangles) correspond to the right axis, respectively.
    }
\label{F2}
\end{figure*}

TRSB can be effectively studied by zero-field (ZF) \msr, which detects the magnetic field inside the bulk of samples with high sensitivity in the absence of external field~\cite{Luke1998, Matsuura2023}. In the TRSB superconducting state, a small magnetic field can be induced near defects or boundaries of the TRSB domains~\cite{Matsumoto1999}, which enhances the relaxation rate of the ZF-\msr\ asymmetry spectrum. Figures\,\ref{F2}a–d, show the results of ZF-\msr\ measured for the collections of single crystals of \FST\ in the nematic phase ($x \sim0.35$) and tetragonal phase ($x = 0.64$, 0.75 and 0.83).
For $x \sim 0.35$ ($T_{\rm c} \sim8$\,K) and $x = 0.64$ ($T_{\rm c} = 14.5$\,K), the ZF-\msr\ asymmetry spectra in the superconducting state ($T < T_{\rm c}$) shows faster relaxation than that in the normal state ($T>T_{\rm c}$) (Figs.\,\ref{F2}a, b). We fit these asymmetry spectra using the stretched and simple exponential functions (as described in Supplementary Information). The temperature dependence of the relaxation rate obtained from the fittings is shown in Figs.\,\ref{F2}e, f. The temperature-independent relaxation observed in the normal state can be attributed to randomly aligned nuclear magnetic moments, fluctuating at a rate slower than the \msr\, timescale. An important observation is that the relaxation rates are enhanced in the superconducting state for both non-spin-rotated (NSR) and spin-rotated (SR) configuration modes, indicating the emergence of magnetic fields in the superconducting state. We also find that the temperature dependence of the relaxatio rate in ZF-$\mu$SR is similar to that in transverse field (TF) measurements (see Fig.\,S5), which represents the temperature dependence of the superfluid density (an order parameter of superconductivity).
Moreover, under a longitudinal field (LF) of 100\,G ($x\sim0.35$) or 200\,G ($x=0.64$), we find that the relaxation rate is much suppressed and its temperature dependence becomes weaker than the ZF results (Figs.\,\ref{F2}e, f). Such a suppression of the LF relaxation is due to the muons being decoupled from the internal field by the weak LF, implying that the enhancement of relaxation observed in ZF-\msr\, is due to a static field rather than a dynamic one. These findings indicate that TRS is broken in the superconducting state in the bulk below an onset temperature $T_{\rm TRSB} (\lesssim T_{\rm c})$ in these compositions ($x\sim0.35$ and $x=0.64$) of \FST. 
In $x =$ 0.64,  the onset temperature of TRSB ($T_{\rm TRSB} \sim 11.5\pm1.5$ K, Fig.\,\ref{F2}f) is consistent with the result of the previous Kerr effect measurement\,\cite{Camron2023}.

In comparisons, the asymmetry spectra of ZF- and LF-\msr\ for $x = 0.75$ and 0.83 single crystals, which are close to the antiferromagnetic (AFM) phase of FeTe, are shown in Figs.\,\ref{F2}c, d. In these compositions, we do not observe any clear AFM transitions by DC magnetization measurements (see Supplementary Information). However, the ZF- and LF-\msr\ spectra of these compositions exhibit two major differences from those of $x \sim 0.35$ and $x = 0.64$. First, for $x = 0.75$ and 0.83, we observe two distinct relaxation components in time spectra of asymmetry: a fast relaxation in the initial time region ($t < 1\,\mu$s) followed by another slower relaxation. We fit these spectra using a two-component exponential function (see Supplementary Information) and obtain the temperature dependence of the fast and slow relaxation rates (Figs.\,\ref{F2}g, h). The fast relaxation rate is considerably larger than the slow relaxation, exceeding $\sim 10\,\mu$s$^{-1}$ at the lowest temperature. Another key difference is that neither the fast nor slow relaxation components are suppressed under an application of LF 200\,G, which is at odds with the results in $x\sim0.35$ and $x=0.64$. 
In addition, the increase of both relaxation rates starts from a characteristic temperature $T^*$ higher than $T_{\rm c}$ for both $x = 0.75$ and $x= 0.83$. These results imply that the fundamental causes of both types of relaxation are unrelated to the superconducting state.

To further clarify the effects of AFM order near the FeTe end, we also measure the $x = 0.94$ samples, which show a clear AFM transition at $T_{\rm N}\sim45$\,K in DC magnetization (see Supplementary Information).
Figure\,\ref{F3}a shows the ZF-\msr\ spectra for various temperatures (2\,K $<T\le70$\,K) in $x=0.94$ single crystals. Similar to the results of $x=0.75$ and 0.83, we obtain two-component ZF relaxation curves, characterized by the fast relaxation in the initial time window followed by the slower relaxation in the AFM phase below $T_{\rm N}$. The fast relaxation is vanishing above $\sim T_{\rm N}$. As shown in  Fig.\,\ref{F3}b, the temperature dependence of the relaxation rate obtained from our fitting (see Supplementary Information) clearly shows that the fast component appears below $T_{\rm N}$, and that both fast and slow relaxations are not suppressed in the LF-\msr. 
A distinct difference from the results of $x\le 0.83$ samples is the significant loss of initial asymmetry $A_0$ at low temperatures in ZF- and LF-\msr\, (Figs.\,\ref{F3}a, d). Such a loss of initial fast asymmetry is commonly observed in the presence of the inhomogeneous magnetic phase\,\cite{Sundar2023, Cheung2018}. To estimate the magnetic volume fraction, we also measure the weak transverse field (wTF)-\msr. The amplitude of low-frequency oscillations in wTF is proportional to the paramagnetic volume fraction, and thus we can estimate the magnetic volume fraction from the amplitude loss (see Supplementary Information). Our results of wTF-\msr\, show that the amplitude loss starts from $T_{\rm N}\sim 45$\,K (Figs.\,\ref{F3}c, d), consistent with the ZF- and LF-\msr, and the magnetic volume fraction is estimated to be approximately 90\% at low temperatures (Fig.\,\ref{F3}e).
Similar results of the amplitude loss in the wTF-\msr\ spectra together with the absence of oscillations in ZF-\msr\ have been reported in previous studies of Fe$_{1+y}$Se$_x$Te$_{1-x}$ with $\sim3$\% excess Fe, which exhibit the incommensurate AFM order~\cite{Khasanov2009}.

Our \msr\ measurements in \FST\ reveal significant differences in the microscopic properties of electronic states between $x\le 0.64$ and $x \ge0.75$.
For $x = 0.75$ and 0.83, we have no evidence of long-range AFM order, but we observe the initial fast relaxation at low temperatures somewhat similar to the AFM state. These compositions are close to the AFM ordered phase ($x\gtrsim 0.9$), and thus the observed fast relaxation is likely attributed to short-range AFM correlations of double-stripe type. Indeed, such an initial fast relaxation is often observed in other materials with short-range ordered or inhomogeneous magnetic phases, such as spin glasses, especially near the magnetically ordered phase~\cite{Hiraishi2014, Sundar2023}.
The presence of the short-range AFM in $x = 0.75$ and $0.83$ is also supported by the ZF and LF-\msr\ spectra, which show the temperature dependence of relaxation rate even above $T_{\rm c}=14$ and $11$\,K. This is in sharp contrast with the spectra for $x\sim 0.35$ and $x=0.64$, which show $T$-independent relaxation above $T_{\rm c}\sim 8$\,K and $T_{\rm c}=14.5$\,K, respectively (Figs.\,\ref{F2}e-h). 

The difference between $x\le 0.64$ and $x \ge 0.75$ is further corroborated by the magnetization and specific heat measurements at low temperatures. For $x\ge 0.75$, a finite residual density of state in the superconducting state is resolved, indicative of a fraction of non-superconducting regions (Fig.\,S2). Thus, our results suggest that the short-range AFM order and superconducting phase coexist below $T_{\rm c}$, contributing to the observed fast relaxation for $x \ge 0.75$. In contrast, the residual density of states in the zero-temperature limit for $x\le 0.64$ is almost zero, which indicates that we have no AFM short-range order, consistent with the absence of the fast relaxation in the ZF-\msr.

\begin{figure*}[tbp]
    \includegraphics[width=0.7\linewidth]{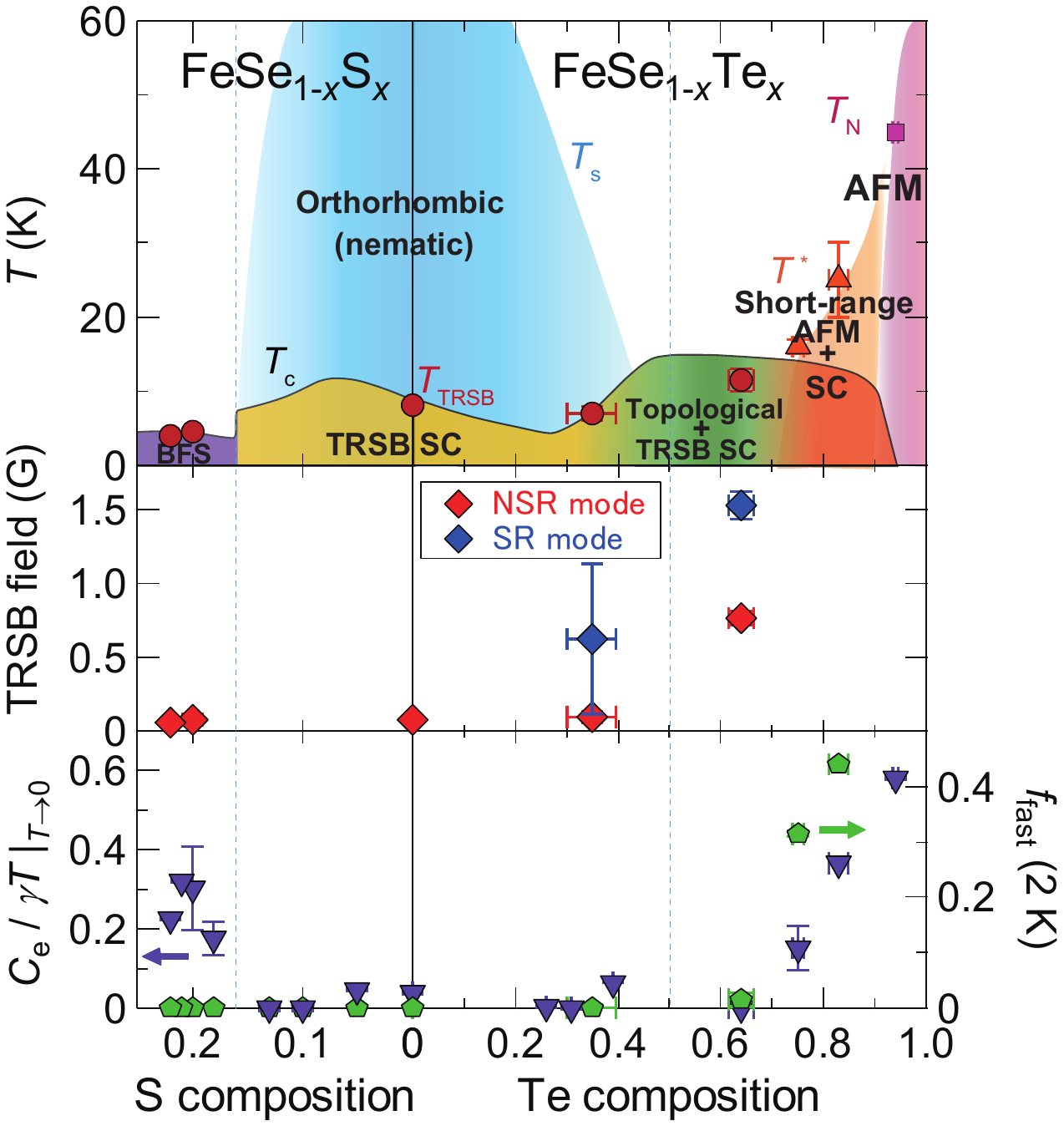}
    \caption{
{\bf Electronic phase diagrams of FeSe-based superconductors.} (Top panel) Schematic phase diagram of temperature versus S/Te composition obtained in this study for \FST, combined with the reported results of \FSS~\cite{Matsuura2023}. 
$T_{\rm s}$ (blue line), $T_{\rm N}$ (pink line), and $T_{\rm c}$ (black line) denote the nematic, antiferromagnetic (AFM), and superconducting transition temperatures, respectively~\cite{Shibauchi2020, Mukasa2021, Ishida2022}. The onsets of TRSB superconducting states $T_{\rm TRSB}$ (red circles) and short-range AFM $T^*$ (orange triangles) are obtained from ZF-\msr. The superconducting state is divided to the possible ultranodal state (purple shade) with Bogoliubov Fermi surface (BFS) in the tetragonal phase of \FSS\ ($x>0.17$), the first superconducting dome inside the nematic phase (yellow shade), topologically nontrivial state with high $T_{\rm c}$ near $x=0.64$ in \FST\ (green shade), and the coexistence region $x\gtrsim 0.75$ between short-range AFM and superconductivity (orange shade).
(Middle panel) Spontaneous field in the bulk of TRSB superconducting states estimated from the NSR (orange diamonds) and SR (blue diamonds) modes of ZF-\msr. 
(Bottom plane) The residual density of states from the electronic specific heat data, $C_{\rm e}/\gamma T$ (purple triangles, left axis), estimated by the extrapolation of $C_{\rm e}/T$ to $T\to 0$, where $\gamma$ is the normal-state Sommerfeld constant. Fraction of fast relaxation in the two-component analysis of ZF-\msr\ at the lowest temperature $\sim2$\,K is also shown (green pentagons, right axis). }
    \label{F4}
\end{figure*}

From our systematic \msr\ measurements of \FST\ together with the previous studies of \FSS~\cite{Matsuura2023}, we have obtained the phase diagram of FeSe-based superconductors as shown in Fig.\,\ref{F4}. Within the nematic phase of pure FeSe, \FSS\ ($x\lesssim0.17$), and \FST\ ($x\lesssim 0.5$), TRSB superconductivity with a small internal field ($\sim 0.1$\,G) is observed, suggesting the realization of the TRSB $s + {\rm e}^{{\rm i}\theta}d$-wave state in the bulk or near the nematic twin boundaries~\cite{Matsuura2023,Shibauchi2020}. In the tetragonal phase of \FSS\, ($x \gtrsim 0.17$), an ultranodal state with the Bogoliubov Fermi surface (BFS) has been proposed \cite{Setty2020}, which can explain the TRSB and the residual density of states observed in clean samples~\cite{Matsuura2023, Sato2018, Hanaguri2018}. In \FST\ side, we find the coexistence of short-range AFM with superconductivity for high Te composition ($x=0.75$ and 0.83), and a long-range AFM order is confirmed for $x=0.94$. Most importantly, we discovered the bulk TRSB superconductivity in the tetragonal phase at $x = 0.64$, without a sign of inclusion of magnetic order. The $x$ dependence of the fraction $f_{\rm fast}$ of the fast relaxation in the ZF-\msr\, spectra and the residual electronic specific heat $C_{\rm e}/\gamma T |_{T\rightarrow 0}$ (left axis on the lower panel of Fig.\,\ref{F4}) indicates the full superconducting volume for $x\le 0.64$. 
In \FST\ in a range including $x = 0.64$, topologically nontrivial electronic structures are found with the TSS, and thus our observation implies that the bulk TRSB superconductivity meets with TSS. This finding clarifies that the bulk TRSB pairing state induces the Dirac gap in TSS and possibly ferromagnetic-like signatures as reported previously~\cite{Zaki2021, Li2021, Camron2023, Mclaughlin2021, Qiu2023}. Our findings provide a fundamental understanding of the complex relationship between the TRSB superconductivity, nematicity, topology, and magnetism in FeSe-based superconductors. 

From our ZF-\msr\ experiments, the TRSB internal field values at the lowest temperature are estimated as $B_{\rm{int}}= 0.09$ (0.62)\,G for $x\sim0.35$ and 0.76 (1.53)\,G for $x =0.64$ in the NSR (SR) mode, respectively.
Within the nematic phase, our previous NSR-mode study for FeSe~\cite{Matsuura2023} reported consistent value of $B_{\rm{int}}=0.1$\,G with the present result for $x\sim0.35$.
FeSe has compensated hole and electron bands, and the gap structure is highly anisotropic, which accounts for the energy proximity of $s$-wave and $d$-wave superconducting states\,\cite{Shibauchi2020}. Such a superconducting state of FeSe can be realized in the complex order parameter $s + {\rm e}^{{\rm i}\theta}d$-wave state with its phase $\theta$ deviating from 0 or $\pi$ in bulk and changes near the twin boundaries, which has been theoretically proposed~\cite{Sigrist1996, Kang2018}. The variation in $\theta$ implies that the superconducting gap has a characteristic change near the twin boundaries, which has been supported in STM and ARPES measurements~\cite{Watashige2015, Hashimoto2018}. 

Meanwhile, in $x= 0.64$, a larger spontaneous field of approximately 1\,G is found in the superconducting state. This value is notably larger than the spontaneous fields typically observed in other TRSB superconductors\,\cite{Luke1998, Matsuura2023}.
As $x= 0.64$ is in the tetragonal phase, the TRSB pairing state with such a large field is different from that in the nematic phase. Indeed, the temperature dependence of electronic specific heat $C_e/T$ at low temperatures shows a very flat dependence with no residual term in  $x= 0.64$ indicative of well-developed full gap, which is in sharp contrast to the strong $T$ dependence with anisotropic gap inside the nematic phase (Fig.\,S2b). A possible TRSB pairing that produces such a large spontaneous field in a tetragonal \FST\ has been proposed in terms of symmetry-based theoretical analysis and the influence of spin-orbit interaction~\cite{Lado2019,Hu2020}. In \FST, the superconducting gap symmetry of $A_{1(2)g}$, $B_{1(2)g}$, and $E_{1(2)g}$ are nearly degenerate, in which case the mean-field theory favors a TRSB pairing with a phase difference of $\pm\frac{\pi}{2}$ between the two degenerate functions~\cite{Hu2020}. Such TRSB pairing can lead to a spontaneous field along the $z$-axis, and the strong spin-orbit coupling with Te can enhance it and induce a Dirac gap below the Fermi energy~\cite{Lado2019,Hu2020}. Our results show a larger relaxation in the SR mode, a configuration more sensitive to the internal field along the $c$-axis, which is consistent with the above theory. However, additional investigations are essential to fully elucidate the nature and detail of the TRSB pairing state in \FST. 

Finally, in the magnetic topological insulator, doping with magnetic elements or making the heterostructure with magnetic materials can break TRS, inducing a Dirac gap. In these systems, tuning the chemical potential within the gap leads to various topological phenomena, including the quantum anomalous Hall effect and topological (axion) electrodynamics\,\cite{Qi2010, Mogi2017, Tokura2019}, making them significant subjects in modern physics.
The present results showing the coexistence of the TRSB superconducting state and nontrivial topology in \FST\ promise a playground for studying novel synergy effects of TRSB and topology in superconductors, with the capability of the chemical potential control by chemical substitution or doping. 
Moreover, our findings provide fundamental information for our understanding of the complex relationship between the superconductivity, nematicity, topology, and magnetism in the iron-chalcogenide superconductors.

\clearpage

\noindent
{\bf METHODS}

\noindent
{\bf Single crystal growth}\\
Single crystals of \FST\,($x\sim$0.35) were grown by the chemical vapor transport (CVT) technique\,\cite{Mukasa2021}. Fe, Se, and Te powders were mixed with AlCl$_3$ and KCl and sealed in a quartz ampoule as transport agents. The temperatures of the source and sink sides were controlled from 620 to 450\,$^\circ$C, respectively. The Te compositions of single crystals were determined by the $c$-axis length measured by X-ray diffraction (XRD)\,\cite{Mukasa2021, Ishida2022}.\\
Single crystals with high Te compositions ($x\ge 0.64$) were grown by the Bridgman method. Initial materials of Fe, Te, and Se with nominal composition Fe$_{1.03}$Se$_{1-x}$Te$_x$ were mixed and sealed into a quartz tube. Obtained quartz ampoule was heated at 1050\,$^\circ$C for 85\,h, and then cooled down to 550\,$^\circ$C at a rate of 4\,$^\circ$C/h under a temperature gradient of around 8\,$^\circ$C/cm, and finally cooled down to room temperature in the furnace. In order to remove excess Fe, as-grown crystals were annealed under Te vapor. They were sealed with pulverized Te into a Pyrex tube and heated at 400\,$^\circ$C for more than 300\,h\,\cite{Koshika2013, Watanabe2020}. The Te compositions of these single crystals were determined by energy-dispersive x-ray (EDX) spectroscopy (see Supplementary Information).

\noindent
{\bf \msr\, experiment and its analysis}\\
\msr\, experiments were carried out at the Centre for Molecular and Materials Science at TRIUMF in Vancouver, Canada, using the LAMPF spectrometer. We performed Zero-field (ZF), longitudinal-field (LF), and transverse-field (TF) \msr\, experiments at the M20D beamline.  For ZF-\msr, to achieve accurate zero fields at the sample position, it was calibrated using a Hall probe and achieved a magnetic field smaller than 1 mG. Both non-spin rotated (NSR) and spin rotated (SR) mode were measured in ZF-\msr. Besides, we performed the LF-\msr\ measurements with NSR mode and TF-\msr\ with SR mode because the magnetic field could be applied along only the $c$-axis of the sample (beam axis).
Samples were aligned with the $c$-axis perpendicular to the beam axis and parallel to the muon spin direction in spin rotation (SR) mode, mounted in wrapped mylar tape. 
More details on the experimental setup and data analysis are provided in Supplementary Information.

\noindent
{\bf Resistivity, magnetization, and heat capacity measurements}\\
The electrical resistivity was measured at ambient pressure by the 4-terminal method using a Physical Property Measurement System (PPMS) from Quantum Design with the lowest temperature of about 2\,K.
The magnetization was measured by using a Magnetic Property Measurement System (MPMS) from Quantum Design.
The specific heat measurements were performed by adiabatic method with using PPMS for $x \ge 0.64$ and the long relaxation method with $^3$He cryostat for $x \sim0.35$\,\cite{Mizukami2023}.

\bigskip

\noindent{\bf Data availablity}\\
\noindent
The data that support the findings of this study are available within the paper and its Supplementary Information. 

\noindent{\bf Acknowledgements}\\
\noindent
We thank M.~V.~De~Toro~Sanchez for technical help and P. Dai, S. Fujimoto, H. Kontani, E.-G. Moon, S. Onari, M. Sato, K. Shiozaki, and Y. Yamakawa for fruitful discussions. 

\bigskip
\noindent{\bf Funding:} This work was supported by Grants-in-Aid for Scientific Research (KAKENHI) (No.\,JP22H00105, JP22KJ1092), Grant-in-Aid for Scientific Research on innovative areas ``Quantum Liquid Crystals'' (No.\,JP19H05824) and Grant-in-Aid for Scientific Research for Transformative Research Areas (A) ``Condensed Conjugation'' (No.\,JP20H05869) from Japan Society for the Promotion of Science (JSPS), and CREST (No.\,JPMJCR19T5) from Japan Science and Technology (JST), and Division Of Materials Research\,(DMR)-2104661 from U.S. National Science Foundation.
G.Q.Z. has been supported in part by CAS Project for Young Scientists in Basic Research (2022YSBR-048).
This work was partially funded by the Max Planck-UBC-UTokyo Centre for Quantum Materials and the Canada First Research Excellence Fund, Quantum Materials and Future Technologies Program.

\bigskip
\noindent{\bf Author Contributions:} Y.J.U. and T.S. conceived the project. M.R., Y.C., K.O., S.L., G.Q.Z., M.O., M.P., C.Y., G.M.L., K.M.K, and Y.J.U. performed \msr\, measurements. M.R., Y.C., K.O., G.Q.Z., K.M., and Y.J.U. analyzed the \msr\, data. M.R., K.O., S.L., K.I., S.F., K.H., and Y.Mizukami performed the resistivity, magnetization, and specific heat measurements. K.O., M.O., T.F., K.M., I.M., D.A.B., T.W., A.Y., and Y.Mizuguchi prepared samples. M.R., Y.J.U., and T.S. prepared the manuscript with inputs from Y.G., G.Q.Z., K.I., K.H., T.F., and T.W. 
All authors discussed the experimental results.

\bibliography{ref.bib}

\clearpage

\clearpage

\clearpage

\begin{figure*}[tbp]
    \includegraphics[width=\linewidth]{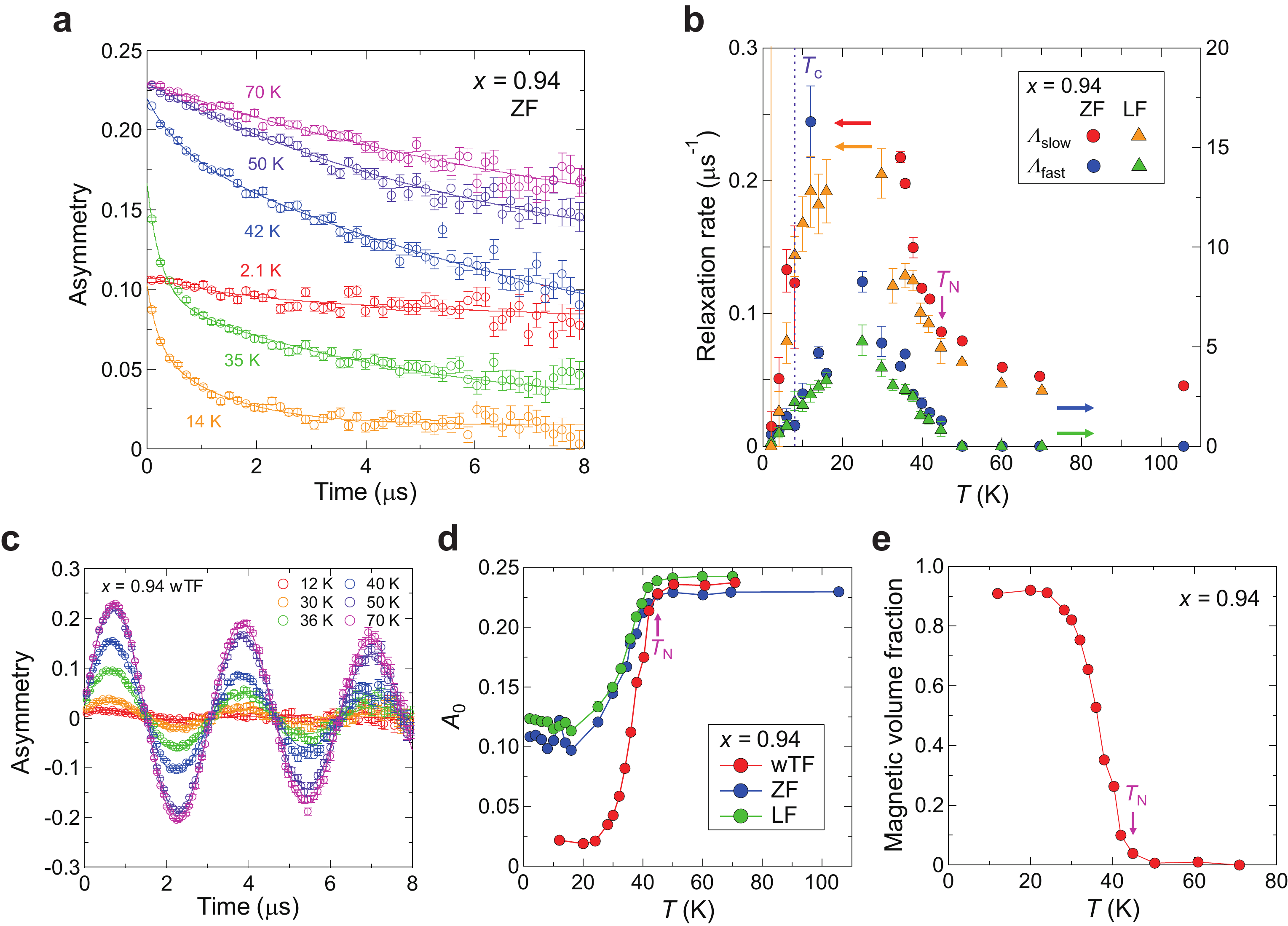}
    \clearpage
    \caption{
       {\bf \msr\ results in FeSe$_{0.06}$Te$_{0.94}$.} {\bf a}, Time spectra of the asymmetry of ZF-\msr\ (NSR mode) for $x= 0.94$ at several representative temperatures. {\bf b}, Temperature dependence of relaxation rate in ZF and LF-\msr\ (NSR mode), obtained by fitting with the two-component function. The symbols for each data and corresponding axes are the same as in Figs.\,\ref{F2}g, h. The onset of the AFM phase $T_N$ is represented by the pink arrow. {\bf c}, Time spectra of the asymmetry in the weak transverse field (wTF)-\msr\ (SR mode) for $x=0.94$ at several representative temperatures. {\bf d}, Temperature dependence of the initial asymmetry $A_0$ in ZF (blue circles), LF (green circles), and wTF-\msr\ (red circles). For wTF-\msr, $A_0$ is estimated from the oscillation amplitude at the initial time. {\bf e}, Temperature dependence of the magnetic volume fraction in estimated from the temperature-dependent loss of the oscillation amplitude in the wTF-\msr\ spectra (see Supplementary Information).
       }
    \label{F3}
\end{figure*}
\clearpage

\end{document}